\documentclass[12pt]{iopart}   

\usepackage{graphicx}
\DeclareGraphicsExtensions{.pdf,.jpg}

\begin{document}

\newcommand{\var}[1]{\mbox{$\mathrm{Var}(#1)$}}
\newcommand{\mean}[1]{\mbox{$\left\langle #1 \right\rangle$}}
\newcommand{\gtwo}[1]{\mbox{$\mathrm{g}^{(2)}(#1)$}}
\newcommand{\musec}{\mbox{$\mu{}$s}}
\newcommand{\mum}{\mbox{$\mu{}$m}}
\newcommand{\muK}{\mbox{$\mu{}$K}}
\newcommand{\ket}[1]{\mbox{$|#1\rangle$}}

\title[An integrated single atom detector]{A single atom detector integrated on an atom chip: fabrication, characterization and application}

\author{D. Heine$^{1,2}$, W. Rohringer$^{1}$, D. Fischer$^{1}$, M. Wilzbach$^{1,2}$, T. Raub$^{1}$, S. Loziczky$^{1}$,  XiYuan Liu$^{3}$, S. Groth$^{2}$, B. Hessmo$^{1,2}$ and J. Schmiedmayer$^{1,2}$}

\address{$^1$ Atominstitut, Technische Universit{\"a}t Wien, Stadionallee 2, 1020 Wien, Austria}
\address{$^2$ Physikalisches Institut, Universit{\"a}t Heidelberg, Philosophenweg 12, 69120 Heidelberg, Germany}
\address{$^3$ Lehrstuhl f{\"u}r Optoelektronik, Universit{\"a}t Mannheim, 68131 Mannheim, Germany}

\ead{schmiedmayer@atomchip.org}

\begin{abstract}
We describe a robust and reliable fluorescence detector for single atoms that is fully integrated into an atom chip. The detector allows spectrally and spatially selective detection of atoms, reaching a single atom detection efficiency of 66~\%. It consists of a tapered lensed single-mode fiber for precise delivery of excitation light and a multi-mode fiber to collect the fluorescence. The fibers are mounted in lithographically defined holding structures on the atom chip. Neutral ${}^{87}$Rb atoms propagating freely in a magnetic guide are detected and the noise of their fluorescence emission is analyzed.  The variance of the photon distribution allows to determine the number of detected photons / atom and from there the atom detection efficiency.  The second order intensity correlation function of the fluorescence shows near-perfect photon anti-bunching and signs of damped Rabi-oscillations. With simple improvements one can boost the detection efficiency to $>$ 95 \%.
\end{abstract}

\vspace{2pc}
\submitto{\NJP special issue: Atom Optics and its Applications}
\maketitle

\tableofcontents
\newpage

\section{Introduction}
Detecting single neutral atoms state selectively is one of the essential ingredients for developing quantum atom optics, atomic physics based quantum technologies and a prerequisite for many quantum information experiments.

The first single atom experiments relied on photo ionization for detection \cite{Hurst:SingleAtomDetection}. Nowadays single atom observation is usually performed optically using fluorescence or absorption detection. Cavity assisted detection schemes are predominant and very successful for absorption detection \cite{Esslinger:AtomLaser,AokiKimble:Microresonator,Kimble:SingleAtomsCavity,Rempe:SingleAtomCavity,MabuchiKimble:cavity,Mab99,Haase:cavity,teper:CavityonChip,trupke:Cavity,ColombeReichel:BECinFibreCavityonChip}, but require active stabilization, increasing the complexity and sensitivity to environmental disturbances.

Fluorescence detection of single atoms can be very efficient if the atom remains localized, since long integration times allow collecting many fluorescence photons. This is realized in single atom in a MOT \cite{Mir03}, dipole traps \cite{Grangier:SingleAtomEmission} or trapped single ions  \cite{Diedrich:AntibunchingfromIon,Blatt:IonTraps,Lei04}. Fluorescence detection of moving atoms which have a short, finite interaction time with the detector is significantly harder and requires both: supreme background suppression and a high collection efficiency to reach single atom sensitivity. For freely falling atoms high collection efficiencies have been achieved using a macroscopic mirror setup covering almost 4$\pi$ \cite{bondo:2006}. Cavities can be used to enhance the fluorescence signal \cite{teper:CavityonChip,trupke:Cavity,Terraciano2009}.

A powerful technique to experiment with ultra cold neutral atoms are atom chips \cite{Fol00} which employ micro-fabricated wires and electrodes to generate magnetic and electric fields for quantum manipulation of neutral atoms few micrometers above the chip surface \cite{Folman2002,Reichel:2002,Fortagh2007}. Many components of integrated matter wave technology have been demonstrated including, combined magnetic/electrostatic traps \cite{Kru03}, motors and shift registers \cite{Han01}, atomic beam splitters \cite{Cas00} the creation of Bose-Einstein condensation (BEC) \cite{Ott01,Hae01,Sch03}, atom interference \cite{Sch05,bohi:09}, the the integration of optical lattices \cite{Gallego2009}.

Recently it was demonstrated that atoms in magnetic guides \cite{Denschlag1999} or on chips can be detected by field ionization near nano-structures \cite{Gruner:09,Gunther:09}, by mounted optical fibers in absorption \cite{Bigelow:FibreDetector04}, in fluorescence \cite{Wilzbach2009}, by using integrated waveguides \cite{Kohnen:2009}, or with the help of cavities \cite{Haase:cavity,teper:CavityonChip,ColombeReichel:BECinFibreCavityonChip}. The latter require active alignment and are technologically very challenging. Miniaturizing the different, highly sophisticated and efficient optics  detection methods benefits from scaling with size \cite{Hor03} and allows integration. 

In this manuscript we describe in detail design, implementation and application of our very simple and robust integrated \emph{fluorescence} detector \cite{Wilzbach2009} that reaches a single atom detection efficiency of $\eta_{\mathrm{at}}=66\%$ and a high signal to noise ratio (SNR) of up to 100 without the need for either localization of the atoms or the assistance of a cavity. The miniaturized detector is based on optical fibers fully integrated on an atom chip \cite{Groth04,Trinker2008} using lithographically fabricated SU-8 mounting structures \cite{liu2005} and allows to examine the statistical distribution of the atoms in a magnetic guide as well as observing non-classical fluorescence emission of single atoms \cite{Heine2009}. In the outlook we discuss a straight forward improvement of our present set-up that will allow us to build atom counters with close to unit efficiency fully integrated on the atom chip.

\section{Basic design}

\begin{figure}[tb]
    \centerline{\includegraphics[width=0.8\textwidth]{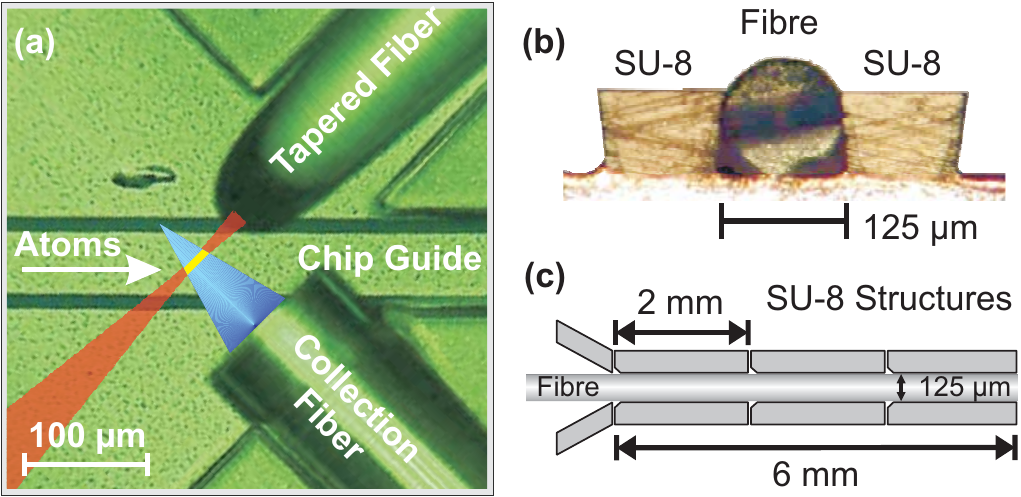}}
    \caption{\label{fig:setupDetector} Details of the integrated atom detector: 
   \textit{(a)} The multi-mode fiber collects light from a cone (blue) determined by its NA.
   The overlap of this cone with the excitation light (red) from the
   tapered fiber defines the detection region (yellow).
   \textit{(b)} Image showing a cross-section illustrating the mounting of the
    optical fibers by the undercut SU-8 structure which clamps the fiber to the chip surface.
   \textit{(c)} Schematic view of the lithographically defined SU-8 holding structure.
   The total length of 6~mm ensures excellent pointing stability,
   while segmentation of into 2~mm long sections allows for thermal
   expansion of the underlying chip. To ease insertion of the fibers
   during construction a funnel has been included in the design.%
    }
\end{figure}

A very simple way of detecting an atom is to observe its fluorescence. The basic idea is to drive a (closed) transition of the atom to an excited state with an external light field and to detect the spontaneously emitted photons.  The sensitivity of the detection and its fidelity depend on the number of scattered photons, the detection efficiency of the scattered light and the suppression of background noise. A well designed fluorescence detector collects as much light as possible and simultaneously has a negligible background. In the ideal case a single detected photon implies that an atom is present in the detection region.

In principle, there is no fundamental limit to the efficiency of a fluorescence detector as long as the atom is not lost from the observation region. Long integration times allows the collecting of many fluorescence photons and efficient discrimination from background. Such a setting was realized for single ions in ion traps \cite{Diedrich:AntibunchingfromIon, Blatt:IonTraps,Lei04}, for a single neutral atom in a dipole trap \cite{Grangier:SingleAtomEmission} or for counting neutral atoms in a magneto-optical trap \cite{Mir03}. Free neutral atoms are considerably harder to detect because the few scattered photons are difficult to distinguish from background light. 

We achieve efficient light collection with small background by employing a fiber optics setup \cite{Wilzbach2009} as shown in figure \ref{fig:setupDetector}(a). Our fluorescence detector is built using a single mode tapered lensed fiber which delivers the excitation light and a multi-mode fiber to collect the fluorescence. The multi-mode fiber is aligned at $90^{\circ}$ angle with respect to the tapered lensed fiber. If no atom is present in the beam focus, very little light is scattered into the fiber. As soon as an atom is present the multi-mode fiber collects the scattered photons.

Such an arrangement is very reminiscent of a confocal microscope. The excitation of the atoms is at a very small focal spot of the lensed fiber.  The collection fiber collects light preferably from the region around the focal spot.  Altogether such an arrangement leads to efficient light collection and superb reduction of background.

\section{Implementation}

For stable and reliable operation both the excitation and detection fibers should be fully integrated into the atom chip. External mountings are poor alternatives since they might move relative to the chip structures and hence lead to non-reproducible results.  The fibers can be glued to the chip by first employing fiber grippers for positioning and alignment. While this method may result in good alignments it is cumbersome to employ and requires significant time and skill to achieve consistent results. Lithographically fabricated robust and precise fiber mounting structures \cite{liu2005} are a great help in assembling the detector. 

\subsection{Fabrication}

Our atom chip holding the micro structures and the detector is fabricated on a $700\,\mu$m thick silicon substrate.  Using optical lithography the micro structures to manipulate the atoms are patterned into an high quality evaporated gold layer \cite{Groth04,Trinker2008} of 2 \mum {} thickness. These wire micro structures support current densities exceeding $2\times 10^7$\,A/cm$^2$ without the risk of wire destruction.

To achieve the precise alignment of the fibers and robust
integration on the atom chip we employ a lithographically
patterned layer of SU-8 on top of the chip structures. SU-8 \cite{delCampo:SU8} is an
epoxy based negative photo resist which is typically developed
using UV-radiation in the range 365-436~nm. SU-8 has high mechanical, chemical and thermal stability. Its
specific properties facilitate the production of thick structures
with very smooth, nearly vertical sidewalls. With a single SU-8
layer a coating film thickness of up to 300~\mum{} is possible.
Once fully developed its glass transition temperature is
approximately $200^{\circ}$C with a degradation threshold around
$380^{\circ}$C. Both temperatures are not reached during operation of the atom chip.

The detection fibers are held in SU-8 trenches 90~\mum{} deep and
125~\mum{} wide, corresponding to the fiber diameter.  Careful
selection of the fabrication parameters produces slightly undercut
sidewalls, as shown in figure \ref{fig:setupDetector}(b). The
layer thickness of 90~\mum{} is chosen to be larger than the fiber
radius, which allows the undercut to effectively clamp the fibers
to the chip surface.

The total length of the trenches was chosen to be 6~mm, sectioned
into three 2~mm long substructures to reduce mechanical stress
during thermal expansion (figure \ref{fig:setupDetector}(c)). The
entry port is funneled to simplify the insertion of the fiber.
The total length guarantees supreme pointing stability, while
allowing accurate positioning of the fibers with longitudinal
alignment precision of a few ten nanometers, as well as excellent
long term stability. Temperature changes and gradients up to
$100^\circ$C resulted in no measurable misalignment of the fibers.
Long term stability of more than three years under experimental
conditions has been achieved.

In fabricating the fiber mounts a $7.5$~\% shrinkage of the SU-8
during development has to be taken into account
\cite{delCampo:SU8}, leading to a slight variability in the
dimensions of the holding structures. It should be noted here that
the variations in fiber diameters are large in comparison.
The diameter of a fiber used in the experiment is
specified with a precision of only 1~\mum{} to a cladding diameter
of $125\pm1$~\mum.  Therefore we fabricated a set of 3 parallel
mounting structures with slightly different sizes (124.5 \mum {},
125 \mum {} and 125.5 \mum ).

To assess the quality and stability of the alignment structures
and the glued fibers, we used the SU-8 structures to hold fiber
optical resonators \cite{FortPhys:2006}. The finesse of the
resonator strongly depends on losses introduced by misalignment.
In our tests we measured a very small additional average loss in
the spliced and mounted resonators of $\sim 0.3$\% which,
neglecting other additional losses, corresponds to a maximal
transversal misalignment of less than $150$ nm.

\begin{figure}[t]
    \centerline{\includegraphics[width=\textwidth]{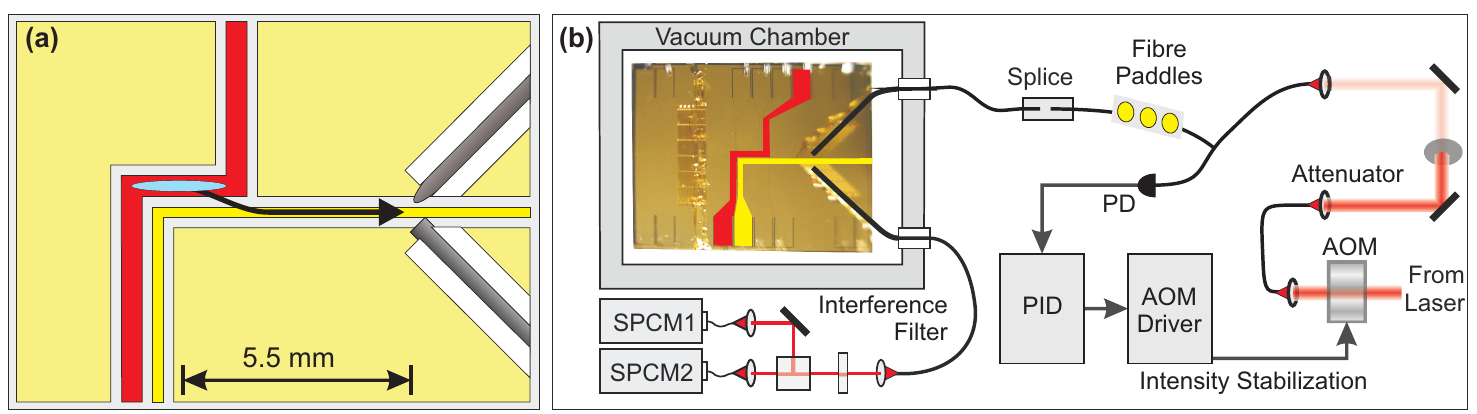}}
    \caption{\label{fig:setupGeneral}
    Details of the atom chip setup and detector: 
    \textit{(a)} Schematic layout of the atom chip.
    Atoms are initially trapped in a magnetic trap generated by a Z-shaped wire (red).
    A magnetic guide (yellow) transports the atoms to the focus of the tapered lensed fiber.
    \textit{(b)} Basic layout of the detector experiment and the atom chip. 
    Light from a frequency stabilized source is delivered via a tapered lensed fiber 
    to the interaction region. Part of the light is split off in a commercial fiber beam splitter and 
    directed to a PIN-type photo diode (PD) for intensity stabilization (PID) using  
    an acousto-optical modulator (AOM) which can also be used as a fast light switch.
    The polarization of the excitation light is adjusted using mechanical fiber paddles, 
    optimizing for maximal fluorescence signal.
    The light scattered by the atoms is collected by a multi-mode fiber that
    guides the light to one single photon counting module (SPCM) for high efficiency
    atom detection or two SPCMs in Hanbury Brown and Twiss configuration for correlation measurements.
    }
\end{figure}

\subsection{Integration in the experiment}

With a diameter of the mounted fibers of 125~\mum, the detection
region is situated 62.5~\mum {} above the chip surface.  To avoid
stray light from the mounting structures when loading the atoms in
the mirror MOT, the detector is built 5.5 mm away from the chip
center. A magnetic guide on the chip transports the atoms to the
detector (see figure~\ref{fig:setupGeneral}).

The fibers mounted on the atom chip are then connected to the
outside of the vacuum chamber via an ultra-high vacuum (UHV)
compatible feed-through \cite{Abraham:TeflonFeedthrough}. The
whole setup can be baked to $100^\circ$ C and achieves UHV below
$10^{-10}$ mbar.

\subsection{Excitation of the atoms}

To excite the atoms to fluorescence the tapered lensed fiber
delivers the light to a focal spot in the guide potential minimum.
Our tapered fibers feature a focal length of $40~\mum{}$ with a
focal diameter of $5$~\mum{}. For the F=2$\rightarrow$F'=3
transition in $^{87}Rb$ saturation intensity is reached in the
focus at a power of $P_\mathrm{sat}=325$~pW. The focus of the
tapered fiber defines the interaction region as shown in figure
\ref{fig:setupDetector}(a) and is located $62.5$~\mum{} (half a fiber
diameter) above the chip surface.

The excitation light is generated by an external cavity diode
laser (ECDL) frequency stabilized to the F=2$\rightarrow$F'=3
transition with a precision better than 1~MHz, smaller than
the natural line width of the transition. The probe laser is
frequency stabilized to a reference laser by means of a frequency
offset (beating) lock. This locking scheme enables us to shift the
lock point by several hundred MHz during operation, allowing to
probe the atomic ensemble over the full range of the Rb-D2
transition.

For additional intensity stabilization part of the light is split
off in a commercial fiber beam splitter and directed to a PIN-type
photo diode (PD) (see figure \ref{fig:setupGeneral}(b)). The PD
signal is fed into custom built proportional-integral-differential
lock electronics (PID) stabilizing the beam power with an
uncertainty below 10~pW to the set value by controlling the output
level of an acousto-optical modulator (AOM) used as a fast light
switch. The polarization of the excitation light can be rotated
employing mechanical fiber paddles.

Using a mechanical fiber splice the stabilized excitation light is
coupled into the single-mode tapered lensed fiber guiding the
light into the vacuum chamber to the detection region.

\subsection{Light Collection}
The light scattered by the atoms is collected by a multi-mode fiber
that guides the light to a single photon counting module (SPCM).
To protect the SPCM and to filter non-resonant background light a
narrow interference filter is included in the beam path (3~nm
FWHM).

The multi-mode fiber has a limited field of view, given by its
numerical aperture of NA$=0.275\pm0.015$ and the core diameter of
$62.5\pm1$~\mum{}. Only light hitting the core at angles up to
maximally $\arcsin(\mathrm{NA})=16^\circ\pm0.9^\circ$ will be
collected. Consequently only emission from atoms within the dark
blue cone marked in figure \ref{fig:setupDetector}(a) is collected
with the full NA. Hence the region of maximum collection
efficiency for the atomic signal is given by the intersection of
the excitation mode field (red) and the full NA cone. This defines
the detection region, which is depicted yellow in figure
\ref{fig:setupDetector}(a).

The height of the detection region is automatically aligned to the
excitation region since the outer diameters of the fibers match.
By aligning the MM fiber such that the collection cone defined by
the numerical aperture overlaps with the excitation region we
reach both a highly selective excitation of the atoms and a
matched collection region. The angle between the tapered
(excitation) and the multi-mode (detection) fiber has been chosen
to be $90^\circ$. This leads to an excellent geometrical
suppression of stray light. Of each nano watt of excitation light
only about 30 photons per second are collected by the detection
fiber, corresponding to a stray light suppression of better than
$10^{-8}$. The detection setup has been positioned such that both
fibers are at $45^\circ$ to the guide to avoid blocking of the
guide.

For our collection fiber (NA=0.275) a maximum photon collection efficiency of
$\eta_{\mathrm{coll}}=1.9$\% can be determined. Together with the
finite efficiency $\eta_{\mathrm{SPCM}}=0.56$ of the SPCM and
optical losses this leads to a total photon detection efficiency
of $\eta_{\mathrm{ph}}=0.9\%$.


\subsection{Experimental Cycle}
In our experiments the atom chip \cite{Fol00,Folman2002} serves as the experimental platform for efficient cooling and trapping of the atoms and for transporting them in a magnetic guide to the detection region.

Our experimental procedure is described in \cite{Wildermuth2004}. The heart
of our setup is a hybrid macroscopic-microscopic atom chip
assembly. It holds the macroscopic wire structures used to
pre-cool and capture the atoms in the primary phase of the
experiment as well as the appropriately designed wire micro
structures on the atom chip needed for trapping and manipulating
the atoms, including the atom detector.

The experimental procedure starts typically with more than $10^8$ $^{87}$Rb
atoms accumulated in a mirror MOT. The atoms are subsequently
optically pumped into the \mbox{$|F=2,m_F=2\rangle$} state and
transferred to a Ioffe-Pritchard type magnetic trap on the atom
chip generated by a Z-shaped wire on the chip surface. The procedure 
is optimized by a genetic algorithm \cite{Rohringer2008}. The atoms
in the magnetic trap are released into a magnetic guide (L-shaped
wire) where the atomic cloud can expand toward the detector. The
position of the magnetic guide above the chip surface is aligned
with the focus of the tapered lensed fiber at the detection region
by adjusting the current through the chip wire and the strength of
the external magnetic field \cite{Folman2002}.

When the atoms are released from the initial Z trap into the
magnetic guide they thermally expand towards the detector. In the
focus of the lensed fiber the atoms are excited by a laser
operating near one of the the rubidium D2 transitions (for example tuned to the $F=2$ to $F'=3$ transition) and the fluorescence photons are collected by the multi-mode fiber and detected by a single photon counting module
(SPCM), their arrival times are recorded with 1~ns resolution. The
experiment is repeated several times to measure the photon
statistics. A typical signal is shown in
Fig.~\ref{fig:varandmean}a.

In the current experiment the phase space density in the magnetic
trap and guide is always less than $10^{-5}$.

\section{Characterization of the atom detector}

\begin{figure}[htb]
    \centerline{\includegraphics[width=\textwidth]{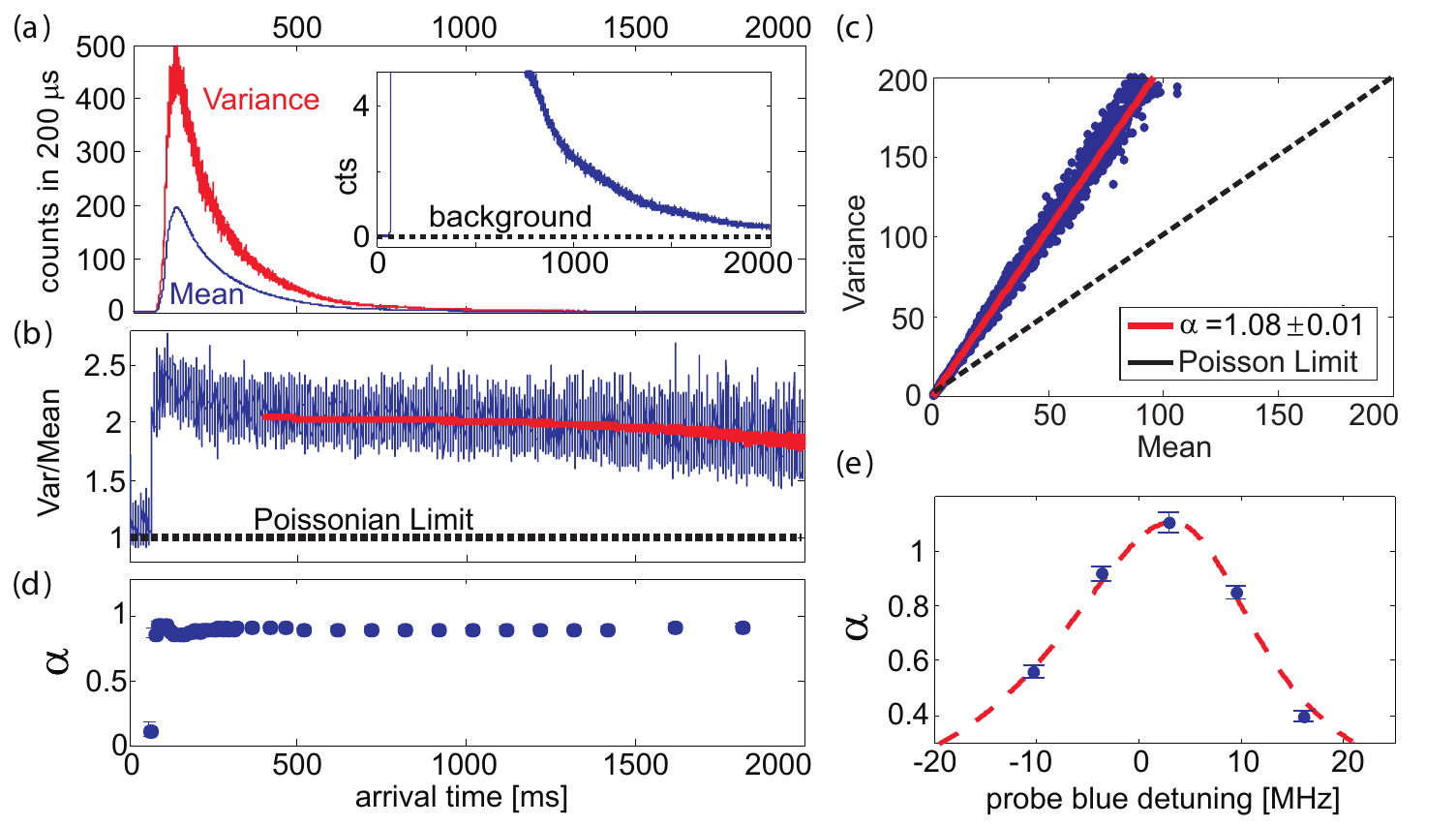}}
    \caption{Photon count signal from the atom detector\\
   \textit{(a)} Mean and variance of the photon counts: 
   The lower (blue) curve shows the mean photon count \mean{n} for 600 experimental runs while the upper (red) curve shows their variance \var{n}. An integration time of $t_{\mathrm{int}}=200~\musec{}$ has been used. Inset: Even after 2~s the signal is still significantly above the background level of 0.06.\\
   \textit{(b)} Ratio \var{n}/\mean{n}: As soon as the atoms arrive, \var{n}/\mean{n} exhibits a sudden jump to a super-poissonian value of $1+\alpha$. Even though $\alpha$ remains constant, the ratio decreases with atomic density as background becomes more important. The red line gives a fit according to equation \ref{eq:varmean1} over more than two orders of magnitude in atom density. Note that the initial overshoot is an artifact of the analysis because around peak the steep sloes can not be neglected during the 200~\musec{} integration time.\\
   \textit{(c)} \var{n} as a function of the mean recorded count rate \mean{n}. An analysis according to equation \ref{eq:varmean1} (red line) yields a signal strength of $\alpha=1.08\pm{}0.01$ counts per atom (integration time $t_{\mathrm{int}}=300$~\musec). The dashed line corresponds to Poissonian statistics.\\
    \textit{(d)} The signal strength $\alpha$ remains constant and is independent of atom density over three orders of magnitude. Here $\alpha$ is evaluated for a shorter  $t_{\mathrm{int}}=50$~\musec{} and shows a reduced value of $\alpha=0.94$ (compare figure \ref{fig:efficiency}).\\
    \textit{(e)} $\alpha$ as function of the probe beam detuning reveals an optimal blue detuning of 3~MHz. The dashed red line is a model based on ref.\cite{bondo:2006}.}
    \label{fig:varandmean}
\end{figure}

One observation in our experiments is that we see the full length of the atom pulse guided in the magnetic guide (Fig.~\ref{fig:varandmean}(a)) even though the resonant excitation light of the detector is always on. These measurements show that in our experiments effects of stray excitation light on the guided atoms can be neglected. This is quite remarkable since magnetic traps are extremely sensitive to light scattering. On average, scattering of a little more than a single photon is sufficient to pump the atom into a magnetically un-trapped state, removing it from the magnetic guide.

Figure \ref{fig:varandmean}(a) shows the atomic signal together with its variance, calculated from 600 consecutive
measurements. The shape can be described by a 1D Maxwell-Boltzmann distribution given by the temperature of the atoms when the reflection of the part of the atom cloud that starts to expands towards the closed end of the guide is also taken into account. The measured temperature of $25~\muK$ agrees with independent measurements via time of flight methods.

\subsection{Background}

The fiber-based detector presented here has exceptionally low
background count rate of only $\sim{}300$ counts per seconds (cps)
at 1nW resonant probe light . The dominating contribution are the dark counts ($\sim{}260$ cps) of the employed photon counter (Perkin-Elmer, SPCM-AQR-12).

Despite the proximity of the atom chip surface (at a distance of
62.5~\mum) the influence of stray probe-light is essentially
eliminated with a suppression of better than $10^{-8}$. 1~nW
excitation light contributes only $\sim{}30$~cps to the
background.

This excellent suppression of stray light is achieved by the fact
that the a tapered lensed fiber delivers excitation light
efficiently to a very small detection region and the multi-mode
fiber, mounted at an angle of 90 degree, selectively collects the
fluorescent photons from this small volume
(Fig.~\ref{fig:setupDetector}a).  Basically all the background
light originating from scattering on the close by chip surface or the
vacuum chamber is blocked very efficiently, because it is not
matched to the guided modes in the collection fiber. In this sense
the setup has characteristics similar to a confocal microscope,
with the difference that the involved point spread functions are
different.

Other background sources are stray light from the outside and the glow of the hot dispensers.  An interference filter efficiently suppresses this remaining background to below 10 cps. (see Fig.~\ref{fig:setupGeneral}b) 

The excellent stray light suppression and overall small background
level allows high fidelity detection of a single atom and accurate
measurements of photon and atom statistics.

\subsection{Signal Strength}
An important parameter of the detector is the total number of photons that can be scattered by the guided atoms before they are lost for the detection process.  We discuss this here for detecting $^{87}Rb$ atoms in F=2 using the F=2$\rightarrow$F'=3 transition.  The total number of scattered photons can be estimated from the ratio of the fluorescence counts when exciting on the closed F=2$\rightarrow$F'=3 transition to the fluorescence counts on the open transition F=2$\rightarrow$F'=1.  Exciting on the later an atom scatters slightly more than one photon before being optically pumped into the other hyperfine ground state, where it remains dark.  We find that that in our detector each atom scatters $\sim 120$ photons when excited on the F=2$\rightarrow$F'=3 transition. 

Combining this number with the overall photon detection efficiency
\mbox{$p_\mathrm{det}=0.9$\%} given above, we estimate that our
detector sees in average $\alpha \sim 1.1$ photon from each atom
passing by. This number was confirmed from independent global atom
number measurements using absorption imaging.

The maximum signal per atom is generated not for resonant illumination, as might be expected, but for a blue detuning of 3~MHz of the exciting light field as shown in figure \ref{fig:varandmean}(e). Looking more into the details, the atoms accumulate a time dependent detuning $\delta{}(t)$ during interaction with the probe field due to photon recoil induced Doppler shift. An initial blue detuning allows pushing the atoms trough resonance to maximize the total number of scattered photons during interaction \cite{bondo:2006}.

\subsection{Photon statistics}

When an atom arrives at the detector, it absorbs and then re-emits photons. A few of these photons are counted by the SPCM. This photon scattering strongly disturbs the atom. Besides the photon recoil it will pump the atom into an un-trapped state and eventually into a different hyper fine ground state which is not excited by the probing light. Consequently after a time $\tau$ the atom will either leave the detection region and / or stop scattering photons.  During detection each atom emits a photon burst, and on average $\alpha$ photons are detected per atom.  These photon bursts lead to a super-poissonian statistics for the detected photons (Fig. \ref{fig:varandmean}(b)). As we will show below we can relate the variance and the mean of the photon counts directly to the average number of photons detected per atom ($\alpha$).

\begin{figure}[t]
    \centerline{\includegraphics[width=0.7 \textwidth]{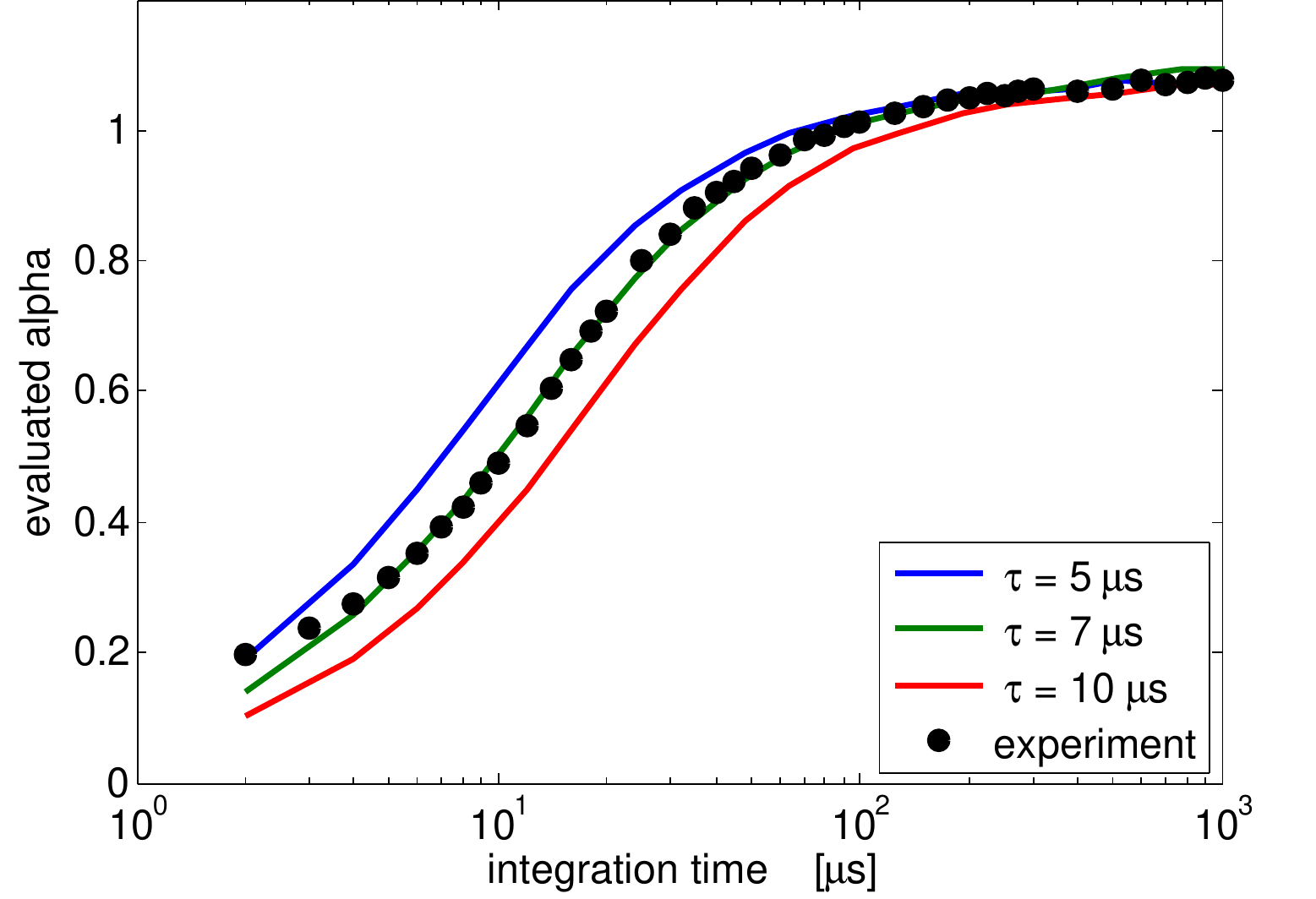}}
    \caption{\label{fig:efficiency} Signal strength $\alpha$ as function of the integration time $t_\mathrm{int}$. While $300~\mu$s are needed to collect the full signal $\alpha$ drops only slowly for shorter integration times.  The measured data is compared to s simple Monte Carlo detection model with different decay rates for the atomic signal.  The model with $\tau=7 \mu s$ gives good agreement. }
\end{figure}

A coherent light source of constant mean photon number \mean{n} creates a Poisson distributed photon stream with variance equal to the mean, $\var{n}=\mean{n}$ and hence
\[
    \left.\frac{\var{n}}{\mean{n}}\right|_{\mathrm{Poisson}}=1.
\]
The superpoissonian photon statistics for the atom detection events can be seen in Fig.  \ref{fig:varandmean}(b) by the jump of \var{n}/\mean{n} $>$ 1 as soon as atoms reach the detector.
The statistical atom number fluctuations lead to increase in the photon noise. The atom number follows its own statistical distribution $p(m)$ around a mean value \mean{m} and Mandel's
formula \cite{RMP:MandelWolf} can be used to describe the
photon count distribution in the presence of atoms. The
probability $p(n)$ to measure $n$ counts is then expressed by
folding the conditional probability
\[
    p(n)|_m = \frac{(m \cdot \alpha)^n}{n!} \, e^{-m\cdot{}\alpha}
\]
to measure $n$ counts in the presence of $m$ atoms with the atom
number probability distribution $p(m,\mean{m})$.
\[
    p(n)=\sum_{m=0}^{\infty}{p(m,\mean{m})\cdot{}p(n)|_m}
\]
The ratio \var{n}/\mean{n} can then be calculated to
\begin{equation}
    \frac{\var{n}}{\mean{n}}=1+ \alpha \, \frac{\var{m}}{\mean{m}}.
    \label{eq:varmean1}
\end{equation}
Hence the ratio of the variance of the photon counts to its mean
value deviates from the Poissonian value 1 by the number of photon
counts per atom $\alpha$ times the ratio of variance to mean of
the atom number.

If in addition a Poissonian background $bg$ with $\var{bg} = bg$
is taken into account equation \ref{eq:varmean1} is modified and
the photon statistics can be expressed as
\begin{equation} \label{eq:varmean2}
    \frac{\var{n}}{\mean{n}}= 1 + \alpha \, \frac{\var{m}}{\mean{m} + \mean{bg}/\alpha}.
\end{equation}

It is worth noting that this relation is independent of the actual
atom number. For a Poissonian atom number distribution,
i.e.~$\var{m}/\mean{m}=1$, the signal strength $\alpha$ can be
measured as the slope of the relation variance versus mean (see
figure \ref{fig:varandmean}(c)). Conversely, if $\alpha$ is known
equation \ref{eq:varmean2} allows examining the statistical
distribution of the atoms by measuring the photon count
distribution. Figures \ref{fig:varandmean}(b) and (c) show the
measured ratio \var{n}/\mean{n} respectively \var{n} versus
\mean{n} together with a fit according to equation
\ref{eq:varmean2}. The signal strength is determined to
$\alpha=1.08\pm0.01$~counts per atom (cpa), independent of atomic
density over three orders of magnitude as demonstrated in figure
\ref{fig:varandmean}(d).

A detailed analysis of the distributions show that an integration time of $300~\musec$ is needed to collect the full signal. Specific applications will benefit from shorter integration
times.  Figure \ref{fig:efficiency}(a) illustrates that $\alpha$ drops
rather slowly when reducing the integration time, while the mean
number of background counts \mean{bg} is a linear function of
$t_{\mathrm{int}}$. For $t_{\mathrm{int}}=100~\musec$ the
integrated signal is $\alpha=1.01$~cpa and even for
$t_{\mathrm{int}}=50~\musec$ we still find $\alpha=0.94$~cpa.

\subsection{Single atom detection efficiency}
\begin{table}[t]
    \centering
    \begin{tabular}{|r|llrr|}
            \hline
            \multicolumn{1}{|c|}{$t_{\mathrm{int}}$} & \multicolumn{1}{c}{$\alpha$ [cpa]}   & \multicolumn{1}{c}{$\eta_{\mathrm{at}}$}  & \multicolumn{1}{c}{$\mathrm{p}_{\mathrm{f}}$}     & \multicolumn{1}{c|}{SNR}\\
            \hline
            $300~\musec$                 & $1.08$                   & $66\%$                                & $9\%$                                         & $11$ \hspace{1pt} $(65)$\\
            $45~\musec$                  & $0.92$                   & $60\%$                                & $1.4\%$                                       & $65$ $(372)$\\
            $20~\musec$                  & $0.72$                   & $50\%$                                & $0.6\%$                                       & $116$ $(655)$\\
            \hline
    \end{tabular}
    \caption{Single atom detection efficiency $\eta_{\mathrm{at}}$ and false detection probability $\mathrm{p}_{\mathrm{f}}$ as function of the integration time for the current system (NA 0.275, $\eta_{\mathrm{ph}}=0.9\%$, $\mean{bg}=310$~cps). Numbers in brackets denote the signal to noise ratio using a low noise SPCM ($\mean{bg}=55$~cps).%
        \label{tab:detefficiency}}
\end{table}

If we require to count at least one photon to see an atom the single atom detection
efficiency is given by 
\[
\eta_{\mathrm{at}}=1-\exp{(-\alpha)}
\]
In our experiment we find $\eta_{\mathrm{at}}=66\%$ for $300~\mu$s integration time.
Background counts during this time lead to a false positive
detection probability of $\mathrm{p}_\mathrm{f}=9\%$. The relation
between detection efficiency $\eta_{\mathrm{at}}$ and integration
time $t_{\mathrm{int}}$ is shown in figure \ref{fig:efficiency}(b)
together with the false positive detection probability for the
current setup and a background reduced setup. While the detection
efficiency is determined by the finite NA of the collection fiber
the SNR is mainly limited by the background. Exchanging the SPCM
by a low noise model with a dark count rate below 25~cps, a total
background as low as 55~cps can be reached.

At $t_\mathrm{int}=45~\musec$ the atom detection efficiency is
$\eta_{\mathrm{at}}=60\%$ with a false detection probability $\mathrm{p}_{\mathrm{f}}$ of only
$\mathrm{p}_{\mathrm{f}}=1.4\%$ (see table \ref{tab:detefficiency}). Defining
$\mathrm{SNR}=\alpha(t_\mathrm{int}) / \left\langle
bg(t_\mathrm{int}) \right\rangle$ a signal to noise ratio
$\mathrm{SNR}=100$ is reached in the current setup at
$t_\mathrm{int}=25~\musec$ ($\eta_{\mathrm{at}}=55\%$). At a
background level of 55~cps $\mathrm{SNR}=100$ will already be
reached at $t_\mathrm{int}=175~\musec$
($\eta_{\mathrm{at}}=65\%$).

\subsection{State Selectivity}
A key signature of an ideal atom detector is the selectivity to internal states, distinguishing between the two hyperfine ground states is most important. Most atoms used in ultra-cold atom experiments have a sizable hyperfine splitting in the ground state, and consequently the two ground states can very easily be distinguished by (near) resonant fluorescence excitation, and state selectivity comes natural.

One characteristic of our fluorescence detector is that for high detection efficiency many scattered photons per atom are necessary, and consequently one needs to use a closed atomic transition. In our present experiment we use $^{87}Rb$ atoms with the two hyperfine ground states (F=1 and F=2 separated by 6.8 GHz).  From the \textit{F=2 ground state} there is a closed transition (F=2$\rightarrow$F'=3) where one can scatter $>$ 1000 photons before changing the hyperfine ground state by off-resonant excitation. 

\begin{figure}[t]
    \center \includegraphics[width=0.69 \textwidth]{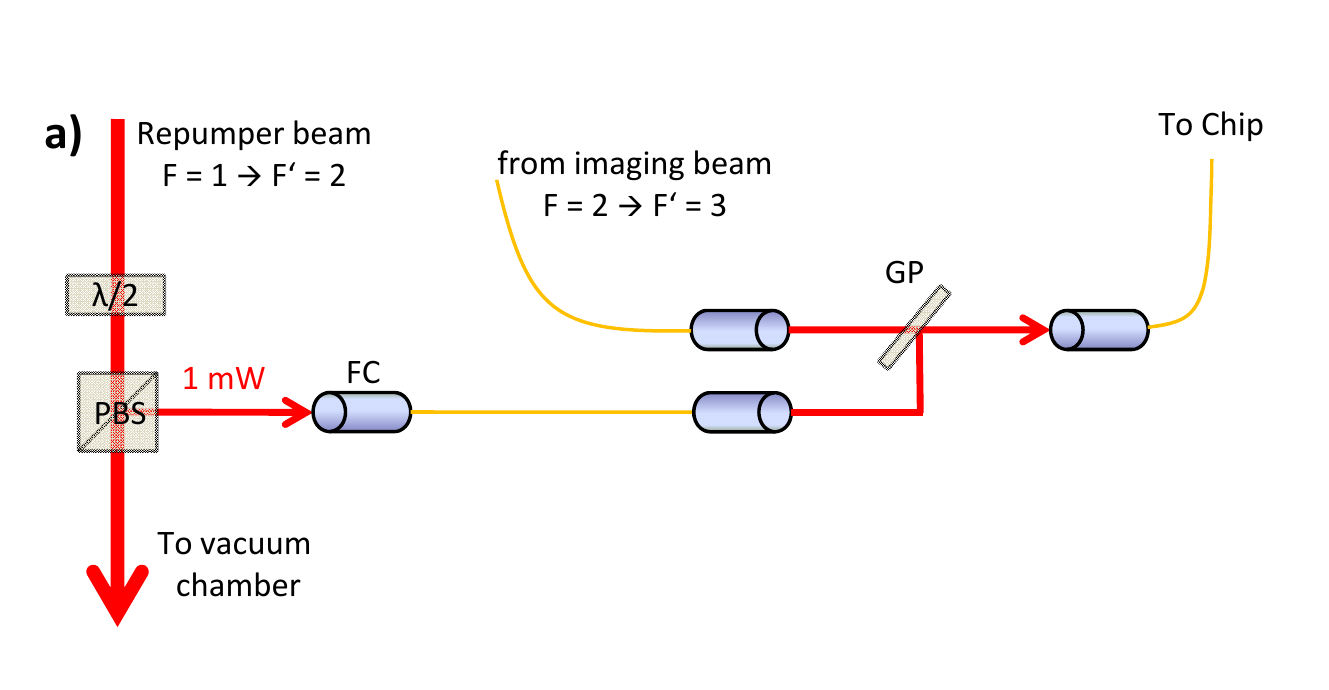} 
    \includegraphics[width=0.3 \textwidth]{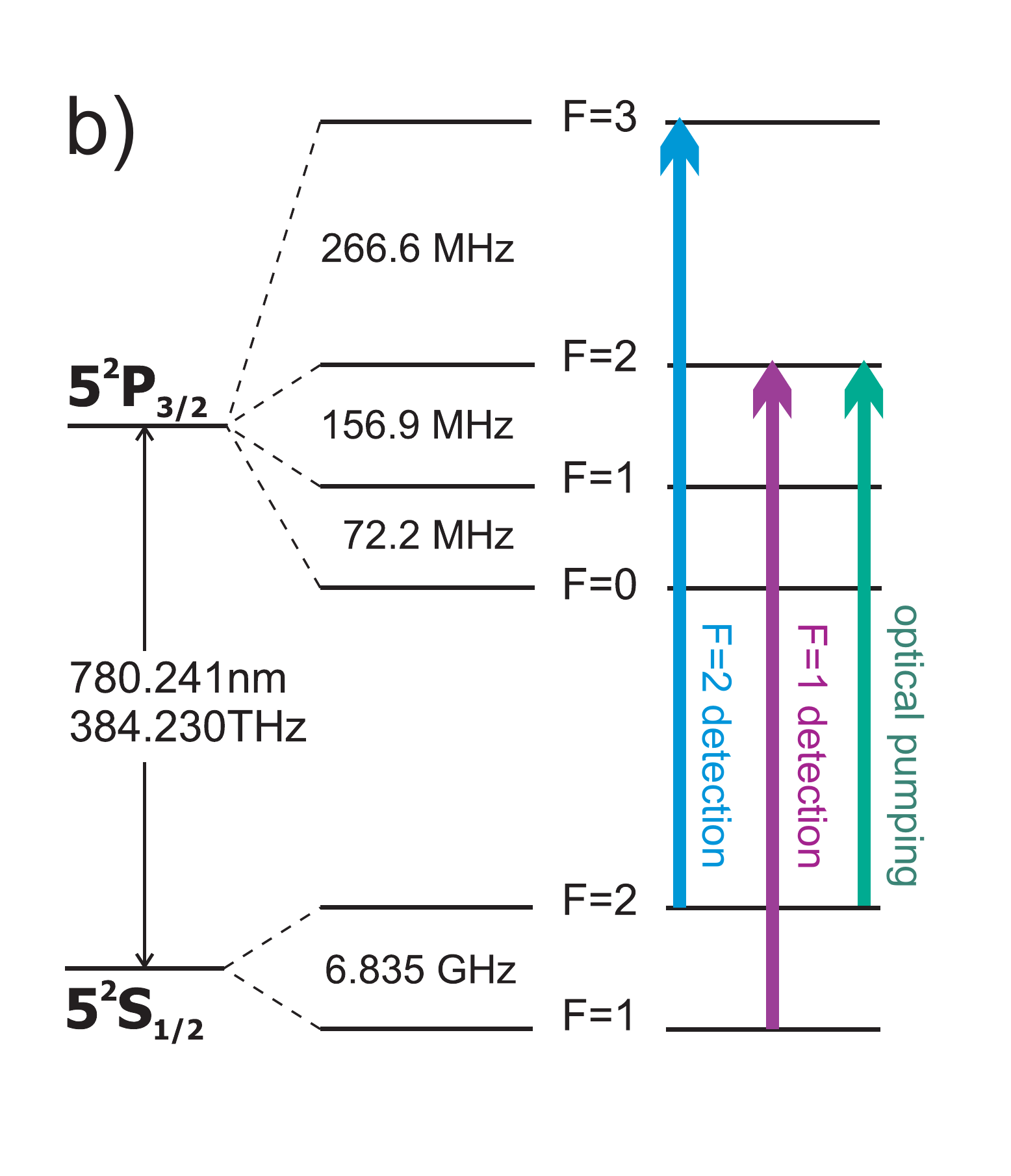} %
    \caption{Changes in the detector setup for F=1 detection: \textit{(a)} Scheme to deliver light to the integrated atom detector for both F=1 and F=2 detection.  The F=1 detection light is taken from the repumping beam of the MOT. \textit{(b)} Level scheme of $^{8}Rb$ D2 line with the relevant transitions for F=2 detection, F=1 detection and the transition used to prepare the F=1 sample by optical pumping from the F=2 to the F=1 ground state. }
    \label{fig:F1Schema}
\end{figure}

For atoms in the \textit{F=1 ground state} the situation is not so simple. To ensure a high scattering rate excitation with F=1$\rightarrow$F'=2 transition is advantageous.  This transition leads to rapid optical pumping into the other hyperfine ground state F=2, and light scattering stops.  This can be overcome by adding a second laser excitation light tuned to F=2$\rightarrow$F'=3.  With both lasers on one can scatter many photons also off atoms in the F=1 hyperfine ground state and detect them. For large intensities ($>$ 300 pW) of light resonant with the F=1$\rightarrow$F'=2 transition, the value of signal strength $\alpha$ has been measured to be comparable to the case of atoms in the F=2 ground state. The results of our measurements are summarized in table \ref{tab:detselectivity}.
\begin{table}[b]
    \centering
    \begin{tabular}{|r|cc|}
            \hline
            \multicolumn{1}{|c|}{} & \multicolumn{1}{c}{Detector F=1}  & \multicolumn{1}{c|}{Detector F=2}\\
            \hline
            \multicolumn{1}{|c|}{Atoms F=1}                 & $1.03$                   & $-$   \\           
            \multicolumn{1}{|c|}{Atoms F=2}                 & $1.03$                   & $1.08$   \\
            \hline
    \end{tabular}
    \caption{Signal strength for detecting F=1 and F=2 Atoms with the F=1 and F=2 detection schemes for an integration time of 300 $\mu$s.  in the F=1 detection schema the F=1$\rightarrow$F'=2 'repumper' light had an intensity of 300 pW. The slightly lower value for the F=1 detection scheme can be associated with excess background due to the additional excitation light. If we turn off the repumper light, atoms in the F=1 ground state do not create a detection signal at all.%
        \label{tab:detselectivity}}
\end{table}

Introducing the re-pumping laser means that a detector that detects F=1 will also detect F=2. For state selective detection this is not so much of a problem, because one can arrange two detectors in such a way that the first detector measures the F=2 atoms.  This detector removes the F=2 atoms from the guide, but leaves the F=1 atoms propagating.  A second detector behind the F=2 detector with both excitation frequencies will register the remaining atoms, all in F=1.

\section{Measurements}

We have employed our detector in a series of experiments to
illustrate its robustness and versatility.

\begin{figure}[t]
    \includegraphics[width=\textwidth]{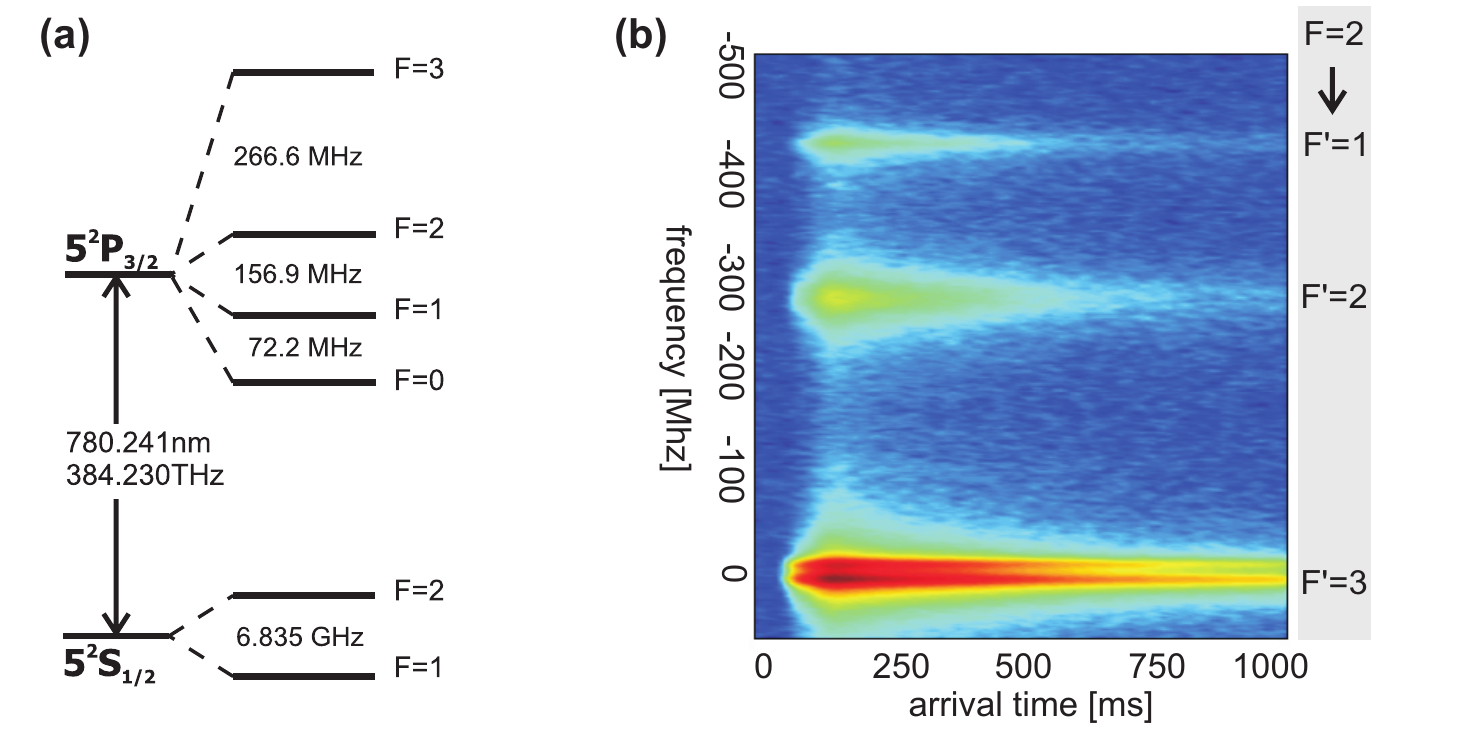}%
    \caption{Spectroscopy of guided atoms: \textit{(a)} Level Scheme of the ${}^{87}$Rb-D2 line. Energy splittings are for zero magnetic field.
    \textit{(b)} Spectroscopy of atoms in the guide by scanning the frequency of the detection light over 600 MHz. Transitions from the $F=2$ ground state to $F'=1,2,3$ excited states are resolved. Note that in the $F'=3$ transition the two cyclic transitions $m_F=2 \rightarrow m_{F'}=3$ and $m_F=-2 \rightarrow m_{F'}=-3$ can be distinguished due to Zeeman splitting induced by the magnetic guiding field minimum which was chosen to be at 4~G for this measurement.}
    \label{fig:spectroscopy}
\end{figure}

\subsection{Spectroscopy in the magnetic guide}
To demonstrate that the integrated fluorescence detector allows
selective addressing of atomic levels we have measured the signal of an atom
pulse passing the detector for different frequencies of the probe
light.

For these measurements the atoms are prepared such that upon arrival at the detector the atoms are in the \ket{F=2,m_F=2}
ground state and can be excited to the $F'=1,2$ or $3$ state.
Figure \ref{fig:spectroscopy} displays the results of a 600~MHz
sweep, starting 60~MHz blue detuned of the $F'=3$ transition and
going down to more than 100~MHz red detuning of the $F'=1$
transition.  The transition F=2$\rightarrow$F'=31, gives
the strongest signal as expected and has correspondingly been
employed for the high efficiency atom detection presented in the
remaining sections of this paper.  The F=2$\rightarrow$F'=1
transition gives the weakest signal, an atom scatters slightly more than one photon before
being optically pumped into the other hyperfine ground state,
where it remains dark. As discussed above the ratio between the signal observed for the  F=2$\rightarrow$F'=X and F=2$\rightarrow$F'=1 transitions can be used to estimate how many photons
the atom scatters at the  F=2$\rightarrow$F'=X transition. In an optimized setting we observe that the $^{87}$Rb atoms scatter about 120 photons on the F=2$\rightarrow$F'=3 transition before they leave the detector.

For the measurement displayed in figure~\ref{fig:spectroscopy}b a
remaining magnetic field at the guide minimum of approximately 4~G
was chosen. This field leads to a Zeeman-splitting of the magnetic
sublevels. The splitting between adjacent levels is smaller than
the line width for the $F'=1$ and $F'=2$ transitions, hence only
one peak is observed. Note though, that for the $F'=3$ transition
the two cyclic transitions
$\ket{F=2,m_F=2}\rightarrow\ket{F'=3,m_{F'}=3}$ and
$\ket{F=2,m_F=-2}\rightarrow\ket{F'=3,m_{F'}=-3}$ can be clearly
separated.  The clear detection signal at the untrapped transition
$\ket{F=2,m_F=-2}\rightarrow\ket{F'=3,m_{F'}=-3}$ demonstrates
that the scattering process is much faster than the dynamics of
atoms leaving the trap.

\subsection{Atom cloud extension in the guide}
\begin{figure}[t]
    \includegraphics[width=\textwidth]{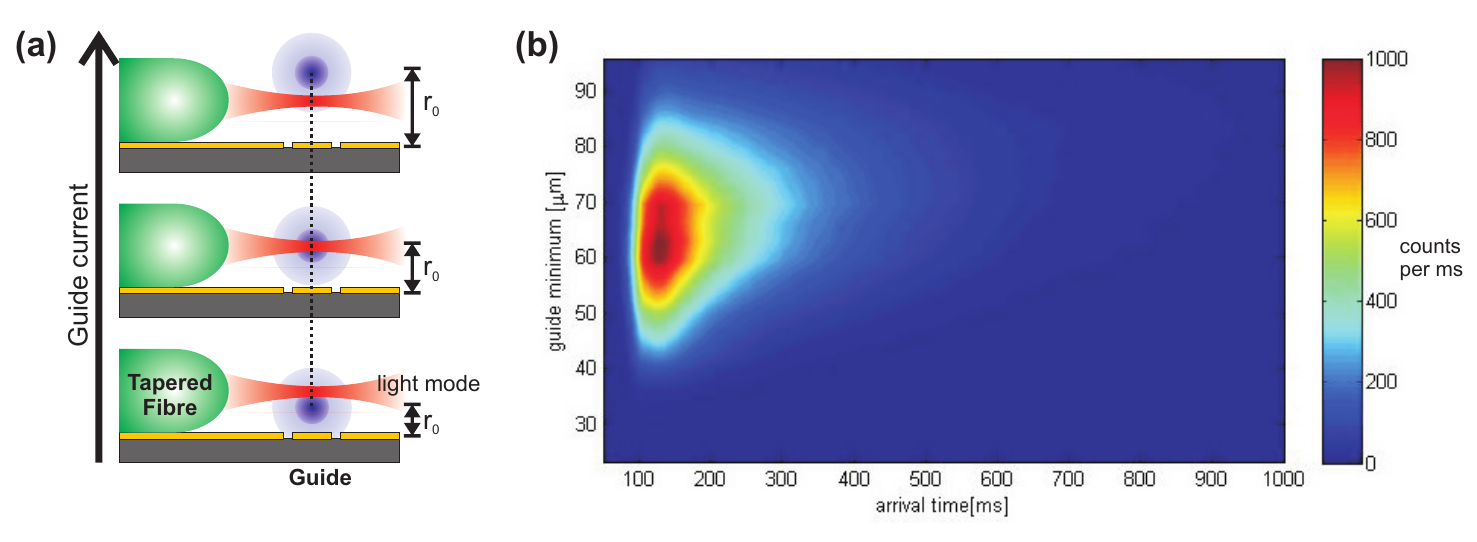}%
    \caption{Probing the profile of atomic density in the guide: 
    \textit{(a)} Scanning the guide current allows probing the guide potential at different heights. A low guide current leads to the guide minimum being close to the chip and atoms propagating higher in the guide are detected. 
    \textit{(b)} The scan reveals the transversal extension of the atoms in the guide potential. The optimal signal is achieved when the guide minimum is aligned with the tapered fiber focus at 62.5 \mum.}
    \label{fig:guidescan}
\end{figure}
The position of the magnetic guide above the chip surface can be
aligned relative to the focus of the tapered lensed fiber at the
detection region by adjusting the current $I_g$ through the chip
wire and the strength of the external magnetic field $B_\perp$
perpendicular to the wire. The guide is located at height
$r_0=(\mu_0/(2\pi)) \cdot{} I_g/B_\perp$, where $\mu_0$ is the
vacuum permeability. Scanning the chip current at constant bias
field, different slices through the guide potential can be probed,
allowing to examine the extension of the atomic cloud in the guide
with only small changes to the potential shape as shown in figure
\ref{fig:guidescan}(a). Figure \ref{fig:guidescan}(b) shows the
result of a scan shifting the guide potential minimum from
25~\mum{} to 95~\mum. The maximum signal is retrieved if the
guide minimum is aligned with the detection area, given by the
5~\mum{} focal spot of the tapered lensed fiber and only the
longitudinally fast atoms reach the outlying regions of the guide.

\subsection{Time interval analysis}
\begin{figure}[t]
    \centerline{\includegraphics[width=0.7\textwidth]{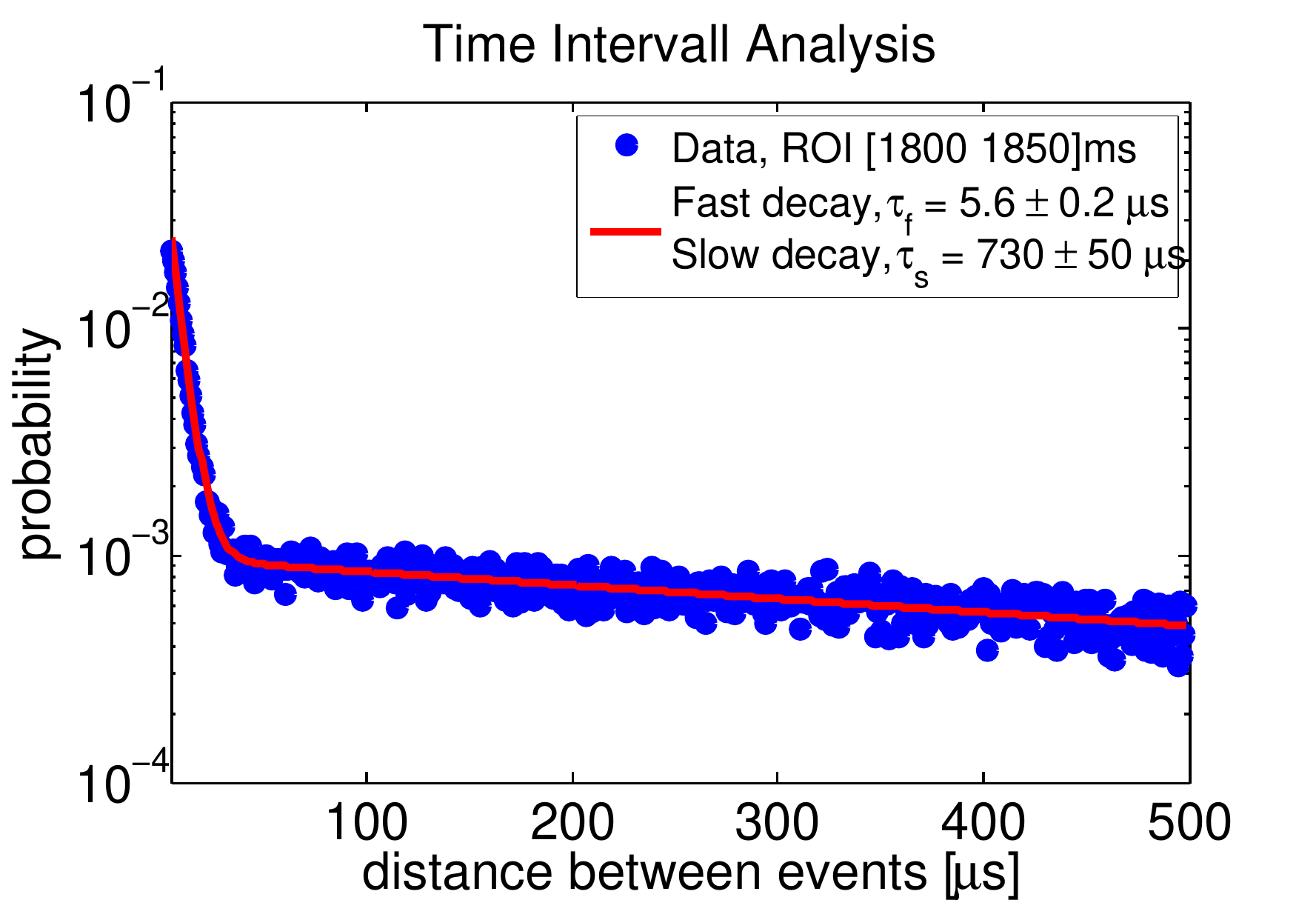}}%
    \caption{Time interval analysis: The data points show the probabilities for finding different time intervals between successive photon detections in the low density tail of the atom distribution. The red line is a double exponential fit described in the text. Short time intervals are dominated by the fluorescence burst of the atoms. The long intervals are given by the atom arrival rate at the detector. An exponential fit demonstrates that the atoms arrive stochastically independent. Additionally a Poissonian distribution of atom arrival times could be confirmed by comparing the slope of the measured distribution to the mean rate of atom detections.}
    \label{fig:TIA}
\end{figure}
%

Time interval analysis (TIA) examines the temporal distance
between neighboring events to draw conclusions on the statistical
distribution of the atoms \cite{EsslingerTIA2007}. In contrast to
the noise analysis presented in section 4.3, TIA does not require a
series of measurements, but extracts statistical information from
a single run of the experiment. Central to TIA is the
probability $p_0$ to measure no counts in a time bin of size $t_b$. 
For example: For independent
counting events, such as multi-mode fluorescence from a constant
source, $p_0$ is a constant and the probability of finding a
time interval of $k$ bins that contain no photon counts is given
by \mbox{$P_{\mathrm{TI}}(k)=(1-p_{0}) \, p_{0}^{k}$}. It follows
that $\log P_{\mathrm{TI}}(k)$ is a linear function of $k$ for
uncorrelated events.

The measured time interval distribution for atom detection events in our detector is shown in
figure~\ref{fig:TIA}.  It is composed of two superimposed exponential
functions (red line, linear on the log scale of the graph). This suggests that we observe
two distinct processes. An atom creates fluorescence counts
according to its instantaneous fluorescence rate while it is
present in the interaction region. After a certain time,
characterized by the $1/e$ interaction time $\tau=12~\musec$, the
atom leaves the detection region. Except for random background
counts with very low probability the detector sees no further
light until the next atom arrives. Therefore, the steep slope for
short time intervals is determined by the instantaneous
fluorescence rate of individual atoms. For long time intervals the
slope is given by the atom arrival rate.

In the experiments presented here we use thermal atoms in a
multi-mode guide (typically $>10^3$ transverse modes are occupied),
thus their statistical distribution can be well approximated by a
Poisson distribution. Indeed the linear relation between $\log
p_{0}$ and $k$ for times $\gg \tau$ shows that the atoms arrive
independently from each other. Additionally $\log p_{0}$ equals the
mean rate of atom detections, verifying a Poissonian distribution
of arrival times in the multi-mode guide.

\subsection{Photon antibunching}
\begin{figure}[t]
    \centerline{\includegraphics[width=\textwidth]{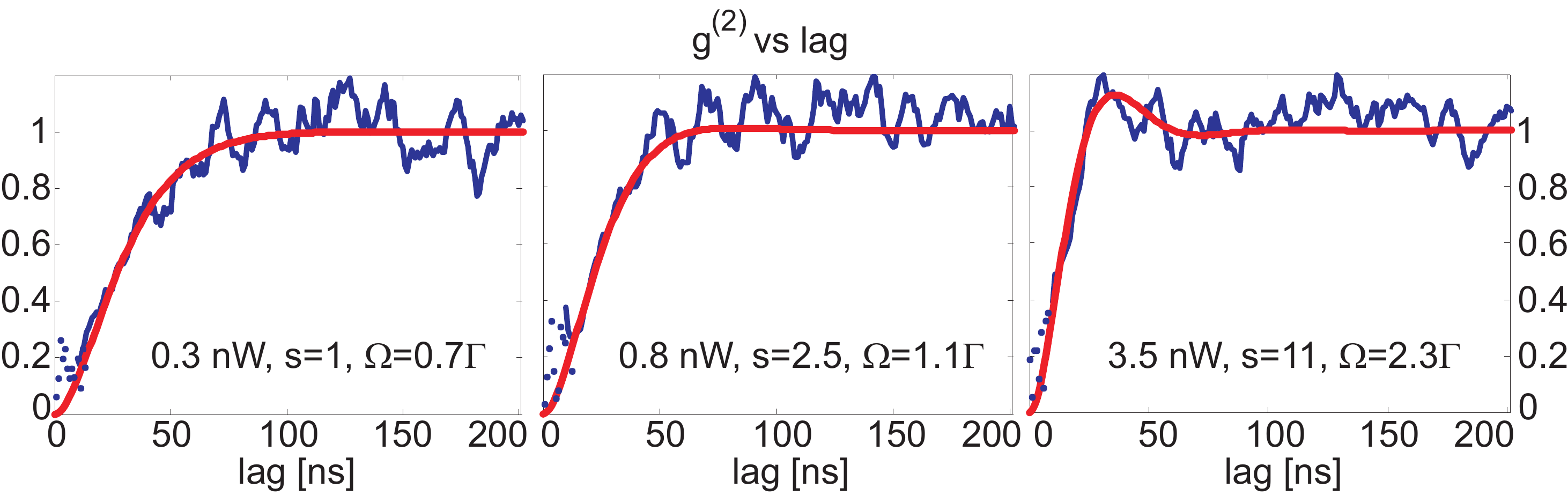}}
    \caption{\label{fig:g2}Photon antibunching:
    Second order intensity correlation function \gtwo{\delta{}t} for Rabi frequencies $\Omega$ increasing from $0.7\Gamma$ ($s=I/I_{\mathrm{sat}}=1$) to $2.3\Gamma$ ($s=11$). For time lag shorter than the excited state lifetime of $1/\Gamma=26.2$~ns the fluorescence exhibits photon antibunching. The blue curve shows the measured correlation function. The red line is a theoretical model (see equation \ref{eq:g2Rabi}) for the corresponding Rabi frequency without free parameters. For small Rabi frequencies \gtwo{\delta{}t} approaches the limiting value of 1 for large lags, while for $\Omega=2.3\Gamma$ a damped Rabi oscillation can be observed. Each graph is the result of 1000 single measurements.
    }
\end{figure}
A single atom can emit only one photon at any given time. Before a
second photon can be emitted, the atom has to be transferred to the
excited state again. Therefore photon antibunching is expected to
occur for single atom detection events at timescales of the order
of the excited state lifetime $1/\Gamma=26$~ns
\cite{Kimble:77,Loudon}. The second-order intensity correlation
function is given by
\begin{equation}
\label{eq:g2}
    \gtwo{\delta{}t}=\frac{\langle \hat{E}^-(t) \, \hat{E}^-(t+\delta{}t) \, \hat{E}^+(t+\delta{}t) \, \hat{E}^+(t) \rangle}{\langle \hat{E}^-(t) \, \hat{E}^+(t) \rangle^2}.
\end{equation}
Photon antibunching is characterized by $\gtwo{\delta{}t}<1$, and
for a single photon $\gtwo{0}=0$ is expected.

For late arrival times larger than 1000~ms the atomic density in the guide is so low that collisional broadening and multi atom effects can be completely
neglected. In this regime an approximation of \gtwo{\delta{}t} for
arbitrary probe beam powers can be given if the illumination is on
resonance \cite{Loudon}.
\begin{equation}    \label{eq:g2Rabi}
    \gtwo{\delta{}t} = 1-\left(\cos{\zeta\,\delta{}t} + \frac{3\Gamma}{2\zeta} \sin{\zeta\,\delta{}t}\right)\,e^{-3/2\,\Gamma\,\delta{}t}
\end{equation}
With $\zeta = \sqrt{\Omega^2-\Gamma^2/4}$ and $\Omega$ denoting the Rabi frequency.

For high efficiency atom detection a single SPCM is employed while
correlation measurements require two SPCMs in Hanbury Brown-Twiss
like configuration.  Consequently for these measurements the light from the collection fiber is sent
on to a 50/50 beam splitter, and both output ports are coupled to
separate SPCMs (see Fig. \ref{fig:setupGeneral}). The counts from the SPCM are recorded and
timestamped at 1 ns resolution by a multichannel counter card
(FAST ComTec P7888 Multiscaler) in a dedicated computer.  Special care has to be taken to compensate for secondary emission effects in the SPCMs.

Figure \ref{fig:g2}(a) shows three different measurements of
$\gtwo{\delta{}t}$ from the cross-correlation of photon counts in
two SPCMs in a Hanbury Brown-Twiss type setup for Rabi frequencies
ranging from $0.7\Gamma$ to $2.3\Gamma$. The correlation function
was reconstructed from the low density tail of the atom
distribution where the mean atomic distance is large enough to
guarantee the presence of at most one atom in the detection region
at any given time. We find beautiful agreement between the
measurements and the theoretically predicted shapes according to
equation \ref{eq:g2Rabi}. For $\Omega=2.3\Gamma$ Rabi oscillations
in the correlation function can be clearly observed. 
We extract a value of $\gtwo{0}=0.05$ from 4400 single measurements at
$\Omega=1.3\Gamma$. The total photon count rate for this
measurement was approximately 3500 cps. If coincidental background
correlations are corrected for the correlation reduces to even
$\gtwo{0}=0.005\pm0.010$, demonstrating that we are able to
observe near-perfect photon antibunching in the emission of single
atoms passing the detector. This is a clear proof that our
detector is capable of detecting single atoms in passage.


\begin{figure}[t]
    \centerline{\includegraphics[width=0.7\textwidth]{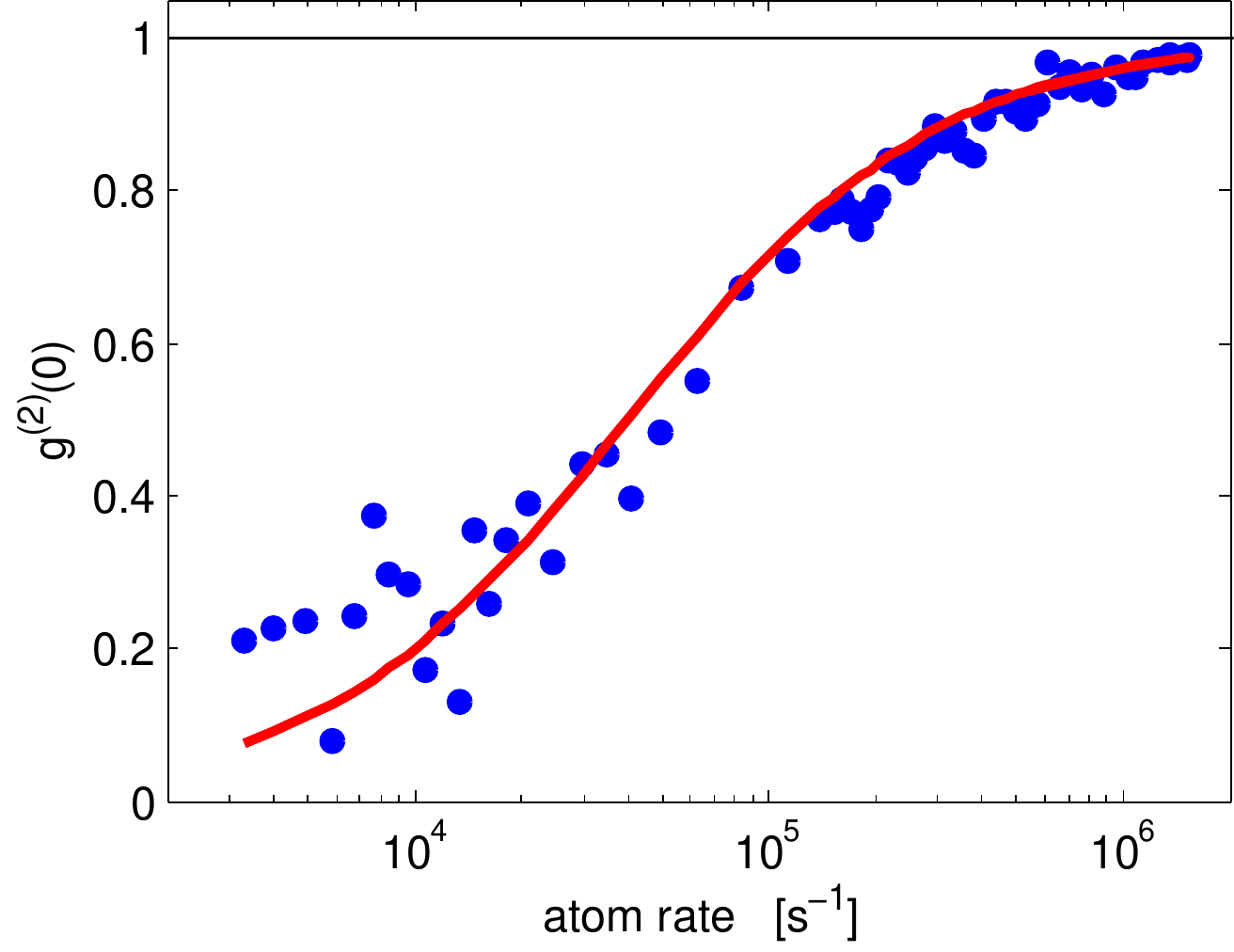}}
    \caption{\label{fig:g2-II}
    Photon antibunching \gtwo{0} as function of the atomic density over three orders of magnitude. The red line is the theoretical limit given by equation \ref{eq:g2fit}.  
}
\end{figure}

Our detector allows us to examine the photon antibunching as
function of the atom density by evaluating \gtwo{0} as function of
the arrival time, as the atom pulse passes the detector (compare
the signal shape in figure \ref{fig:varandmean}(a)). For a
single-mode field with mean photon number \mean{n} the
second-order correlation at lag 0 is limited by \cite{Loudon}
\begin{equation}
    \gtwo{0}\geq{}1-\frac{1}{\langle n \rangle} \qquad  \forall \langle n \rangle \geq 1.
\label{eq:g2n}
\end{equation}
while for $\mean{n} < 1$ the lower limit is 0. For a Fock state
with fixed photon number $n$ the inequality (\ref{eq:g2n}) becomes
an equality and the minimal value of \gtwo{0} is reached. While
for classical light sources $1 \leq{} \gtwo{0} \leq{} \infty$
holds, the region $\gtwo{0} < 1$ is exclusively non-classical and
can only be reached by quantum emitters.

Since $\gtwo{\delta{}t}$ is evaluated when at least one photon
count has been recorded, the mean photon number \mean{n} has to be
calculated under the condition $n \geq{} 1$. This leads to
\mbox{$\mean{n} = \alpha \mean{m} / (1 - \exp{(-\alpha
\mean{m})})$} with mean atom number \mean{m} emitting. Hence \gtwo{0} is
limited by
\begin{equation}
    \gtwo{0}\geq{}1-\frac{1-\exp{(-\alpha \mean{m})}}{\alpha \mean{m}}.
\label{eq:g2fit}
\end{equation}
As can be seen from figure~\ref{fig:g2-II} the measured \gtwo{0}
follows the expected shape for the full atom pulse duration. With
this measurement we extend the original experimental investigation
\cite{Kimble:77,Kimble:78} of the influence of atomic density on
the second-order correlation function over almost three orders of
magnitude change in atomic density.

\section{Future improvements}
The detection efficiency $\eta_{\mathrm{at}}=66\%$  of the current
detector setup is mainly limited by the photon collection
efficiency $\eta_{\mathrm{coll}}=1.9\%$ given by the numerical
aperture NA=$0.275$ of the multi-mode collection fiber. The
detection efficiency can be significantly improved by increasing
$\eta_{\mathrm{coll}}$.

\begin{figure}[tb]
    \centerline{\includegraphics[width=0.5\textwidth]{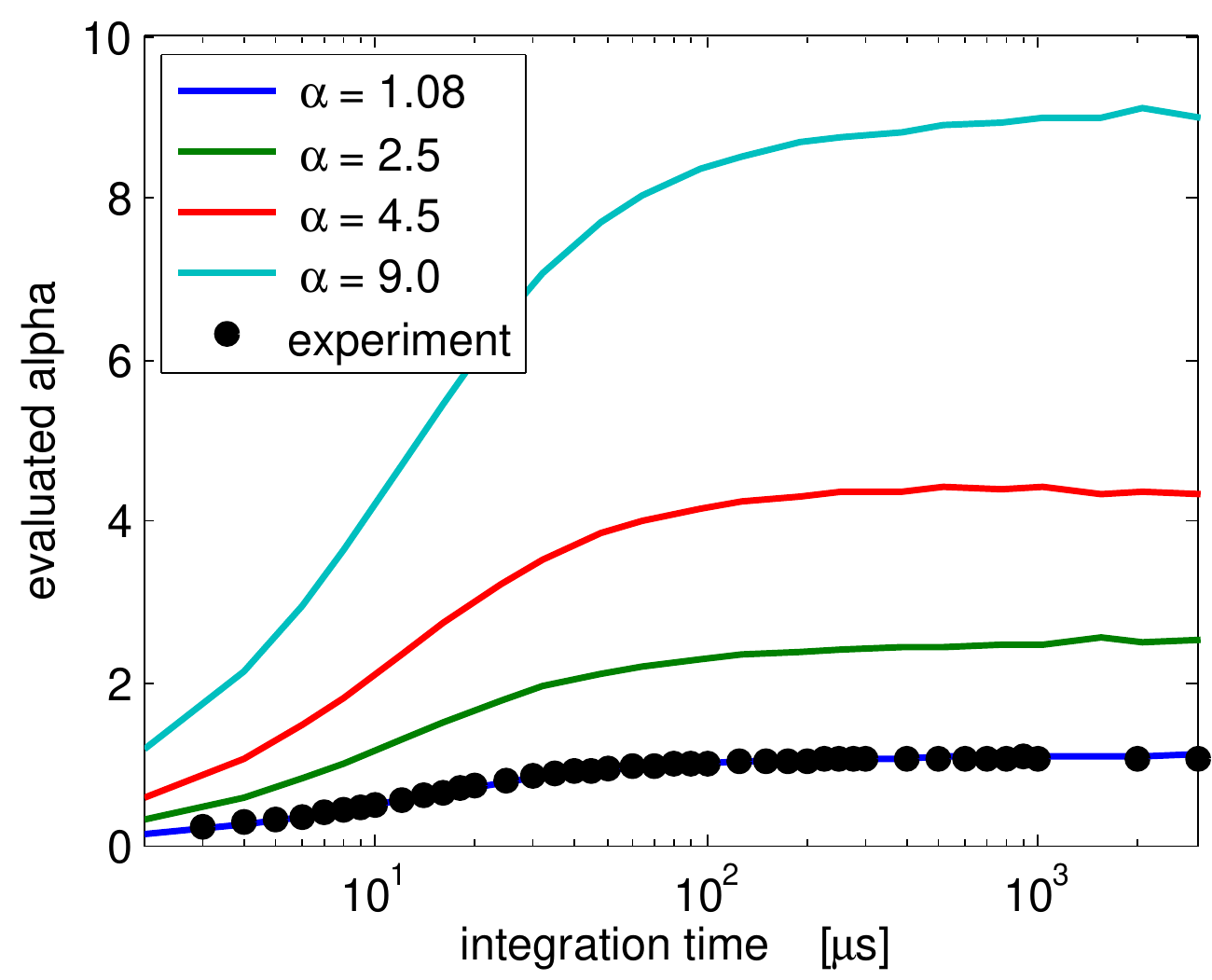},
    \includegraphics[width=0.5\textwidth]{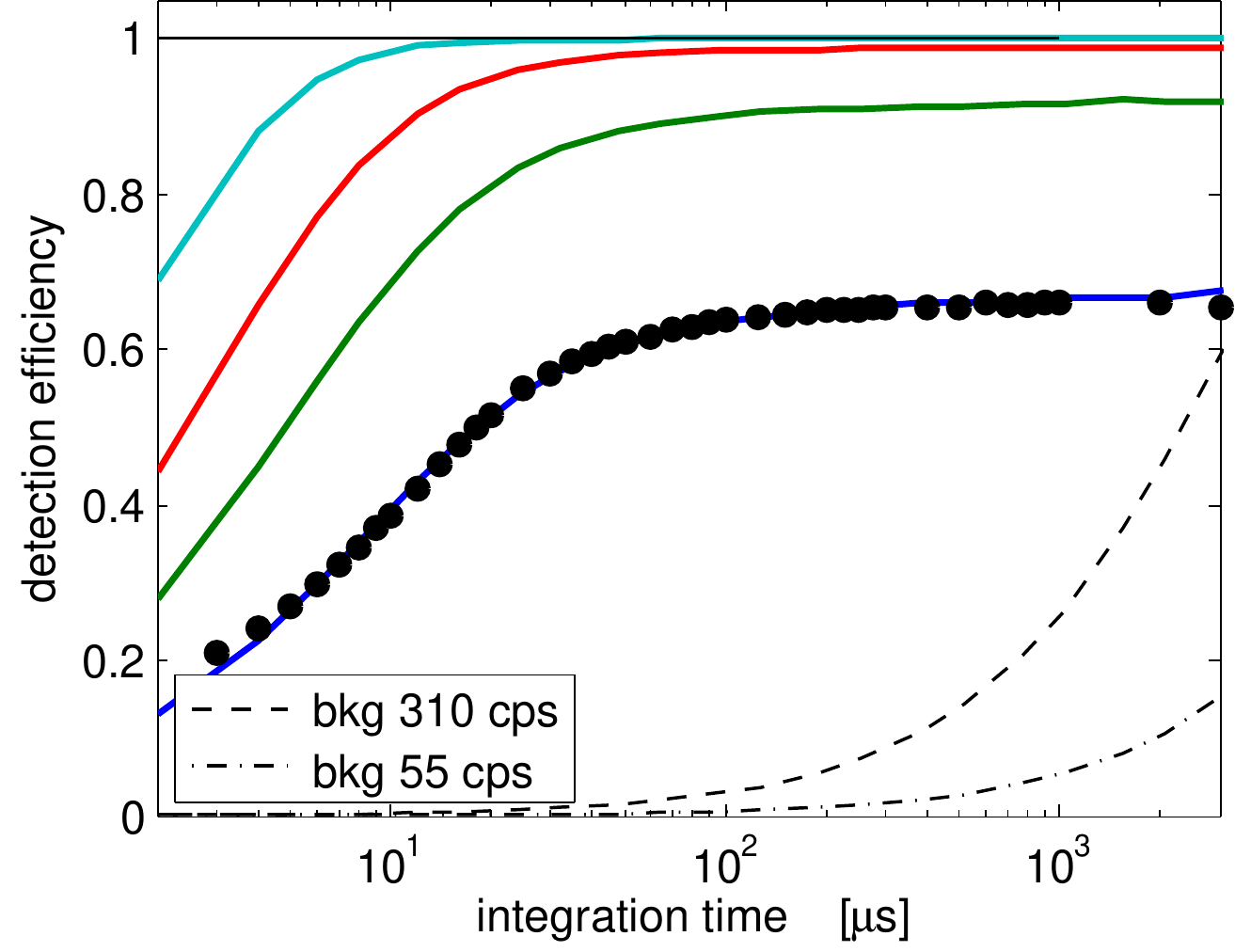}}
    \caption{\label{fig:ImprovedDetector} Projected performance of the detector when light collection is improved to yield $\alpha = $ 2.5, 4.5 or 9.0 counts per atom respectively.  The performance of the present detector with $\alpha = 1.08$ is shown for comparison.
   \textit{(left)} $\alpha$ vs integration time. 
   \textit{(right)} Projected detection efficiency compared to the false detection probability for the two different background levels discussed in the text. %
    }
\end{figure}
A single atom detection efficiency close to unity can be achieved
by substituting the present detection fiber (NA=0.27) by a
fiber with NA=$0.53$. This increases the photon collection
efficiency to $\eta_{\mathrm{coll}}=7.6$\%, which would result in 
$\alpha=4.5$ counts per atom (see table
\ref{tab:detefficiency-imp}).  Such an improved system is expected
to reach even at 50~kHz bandwidth a single atom detection
efficiency of $95\%$.

The detection efficiency can be further improved by collecting
light from two sides by mounting two facing high NA detection
fibers.  Using two NA=0.53 fibers one can reach
$\eta_{\mathrm{at}}>99\%$ at integration times below
$t_{\mathrm{int}}=20~\musec$. For such a detector system a signal
strength of $\alpha=9$~cpa is expected.

With either improvement implemented, high fidelity atom counting
(in the sense of identifying the number of simultaneously present
atoms) and observation of atom transits in transient count rate
analysis will become feasible.

\begin{table}[h]
    \centering
    \begin{tabular}{|r|llrr|}
            \hline
            \multicolumn{1}{|c|}{$t_{\mathrm{int}}$} & \multicolumn{1}{c}{$\alpha$ [cpa]}   & \multicolumn{1}{c}{$\eta_{\mathrm{at}}$}  & \multicolumn{1}{c}{$\mathrm{p}_{\mathrm{f}}$}     & \multicolumn{1}{c|}{SNR}\\
            \hline
            $300~\musec$                 & $4.5$ $(9.0)$    & $99\%$ $(>99.99\%)$       & $<4.3\%$ $(<7.6\%)$               & $273$ $(545)$\\
            $45~\musec$                  & $3.8$ $(7.6)$    & $98\%$ $(>99.9\%)$        & $<0.7\%$ $(<1.2\%)$               & $230$ $(460)$\\
            $20~\musec$                  & $3.0$ $(6.0)$    & $95\%$ $(>99.7\%)$        & $<0.3\%$ $(<0.5\%)$               & $182$ $(364)$\\
            \hline
    \end{tabular}
    \caption{Single atom detection efficiency as function of the integration time for an improved system using a commercially available fiber with NA 0.53 ($\eta_{\mathrm{ph}}=3.8\%$). Numbers in brackets denote the corresponding values for two detection fibers of NA 0.53 ($\eta_{\mathrm{ph}}=7.6\%$). The fale detection probablilty $\mathrm{p}_{\mathrm{f}}$ and the signal to nois ratio (SNR) are given for the present background rate of 310 cps and the improved rate of 55 cps for low dark count photon counting modules.
        \label{tab:detefficiency-imp}}
\end{table}

\section{Conclusion}
To conclude, we have built and evaluated an optical fiber-based
atom detector which is fully integrated on an atom chip, alignment
free by fabrication, and mechanically very robust and capable of state selective
single atom detection at an efficiency of
$\eta_{\mathrm{at}}=66\%$ by counting fluorescence photons. It
enables spatially and spectral highly selective probing while at
the same time offering extremely robust operation.  The total
background is dominated by the dark count rate of the employed
SPCMs. High efficiency detection is possible as well as low
noise, high bandwidth detection. At 20~\musec{} integration time
$\mathrm{SNR}>100$ single atom detection can be performed at an
efficiency of $\eta_{\mathrm{at}}=50\%$. The disadvantage of
fluorescence detection is the destructive nature of the process, which our detector shares with photon detection.
Our detector can therefore be best characterized as the atom optic equivalent of an avalanche
photo diode type SPCM.

Low noise, high efficiency and insensitivity to stray light is
achieved using fiber optics to create very selective excitation of
the atoms in a small, matched observation volume.

The detection efficiency is currently limited by the numerical
aperture of the multimode collection fiber. A straightforward
substitution of the employed NA=0.275 fiber by a commercially
available fiber with NA=0.53 increases the photon collection to
$\alpha$ = 4.5 counts/atom and the single atom detection
efficiency to 95\% at 50 kHz bandwidth. Employing two collection
fibers $\alpha$ = 9 cpa and $\eta_{\mathrm{at}}>99\%$ at
integration times below $t_{\mathrm{int}}=20~\musec$ can be
achieved. With these improvements, full atom counting becomes
feasible.

The high efficiency, signal-to-noise ratio and bandwidth of our
integrated detector make it suitable for many physical systems
where only a few photons can be scattered.  A prime example would
be detecting trapped cold molecules. With its extremely low
sensitivity to stray light, our detector is well suited for
studies of correlated atomic systems and scalable quantum
experiments on a single-atom or molecule level.

\ack
We thank A.~Haase, and M.~Schwarz for help in the early stages of the experiment, and K.-H.~Brenner, and I.~Bar-Joseph for help in the fabrication. We gratefully acknowledge financial support from the Landesstiftung Baden-W{\"u}rttemberg, the European Union (HIP, CHIMONO), the Austrian nano-initiative (PLATON-NAP) and the FWF.  The atom chip shown in this work was fabricated at the The Braun Submicron Center at Weizmann Institute of Science, Rehovot, IL.

\newpage

\section*{References}

\end{document}